# PneumoXttention: A CNN compensating for Human Fallibility when Detecting Pneumonia through CXR images with Attention

**Sanskriti Singh (9th grade, Basis Independent Silicon Valley)**

## Abstract

Automatic Chest Radiograph X-ray (CXR) interpretation by machines is an important research topic of Artificial Intelligence. As part of my journey through the California Science Fair, I have developed an algorithm that can detect pneumonia from a CXR image to compensate for human fallibility. My algorithm, PneumoXttention, is an ensemble of two 13 layer convolutional neural network trained on the RSNA dataset (RSNA, 2020), a dataset provided by the Radiological Society of North America, containing 26,684 frontal X-ray images split into the categories of pneumonia and no pneumonia. The dataset was annotated by many professional radiologists in North America. It achieved an impressive F1 score [0.82] on the test set (20% random split of RSNA dataset) and completely compensated Human Radiologists on a random set of 25 test images drawn from RSNA and NIH (NIH, 2017). I don't have a direct comparison but Stanford's Chexnet has a F1 score of 0.435 on the NIH dataset for category Pneumonia.

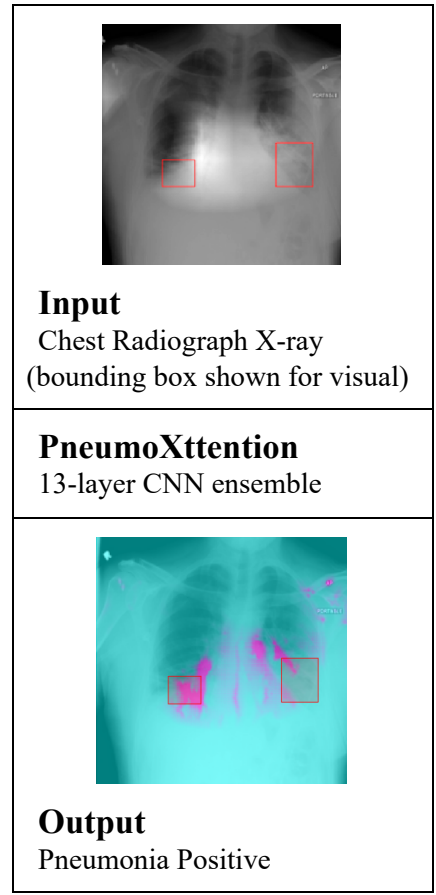

**Input**
Chest Radiograph X-ray
(bounding box shown for visual)

**PneumoXttention**
13-layer CNN ensemble

**Output**
Pneumonia Positive

*Figure 1:* PneumoXttention is an ensemble of two 13-layer convolutional neural networks that takes a chest x-ray image as its input, and outputs whether or not the image has pneumonia. In this example, PneumoXttention accurately detects pneumonia in the image while looking in the correct areas to detect it.

## 1. Introduction

Pneumonia is one of the deadliest diseases for children age 5 and below. Pneumonia accounts for 16% of all deaths of children under 5 years old, killing 920,136 children in 2015 (WHO, 2017). In the US alone, there have been 48,632 deaths due to pneumonia in the year of 2016 (FastStats - Pneumonia, 2017). Additionally, each year in the United States, more than 250,000 people have to seek care in a hospital due to pneumonia. Unfortunately, about 50,000 people die from the disease each year in the United States alone (CDC, 2019). As of today, Chest Radiograph X-ray (CXR) is the best method for diagnosing pneumonia since by looking at the visible inflammation or opacity in the xray pneumonia can be easily identified (NIH, 2018). When analyzing CXR images, most radiologists are trained to look at "review areas" which refer to the areas where there is often a missed finding, or where attention should be given. Often radiologists can accurately predict pneumonia when the opacity is clear and visible, but during times when pneumonia appears in areas such as near the apex or behind the heart, there is a higher chance of misdiagnosis (Manickam). I wanted to train a model that can do better in these areas and hence compensating for human fallibility. Our model, PneumoXttention, is an ensemble of two 13-layer convolutional neural networks that takes a CXR or chest radiograph image as its input and returns whether pneumonia is visible in the image (shown in Figure 1). The models were trained on the RSNA dataset, which contains 26,684 frontal-view chest X-ray images, all of which were labeled as 0 or 1 (RSNA, 2020). 0 meaning no pneumonia and 1 meaning the image has pneumonia. Pneumonia can be hard to detect for a number of reasons, one of which is that the opacity can be easily hidden in certain areas of the image. Pneumonia appears as opacity on the CXR and detecting it on top of whitened areas is difficult for humans. It can also be easily mistaken for other abnormalities which are also diagnosed through x-ray and appear as opacity on the CXR (Manickam). PneumoXttention can provide a valuable second opinion which will result in a higher diagnosis accuracy. I want to create a model which will try to detect Pneumonia as best as it can but I want to compare it with a human Radiologist to see if it makes different mistakes or the same. In case of difference in results, a Human Radiologist can pay more attention and take a third opinion in diagnosis, overall significantly reducing the misdiagnosis and saving lives.

|  | F1 score (25 images) |
|---|---|
| Normal CNN without attention | **0.70** |
| CNN with a unique form of attention | **0.9** |
| Model and Radiologist working together | **1.0** |

*Table 1*: In the table above I am comparing my final algorithm with the unique form of attention to one without any attention. I am also showing the final f1 score of a model and human working together achieving 100%.

## 2. PneumoXttention

### 2.1 Problem Formation

The detection of pneumonia is a binary classification problem. The input of this model is a frontal view of a CXR, and the output of the model is either 0 or 1, 0 meaning no pneumonia and 1 meaning pneumonia. I used binary cross-entropy loss to train the model.

### 2.2 Model Architecture and Training

PneumoXttention is an ensemble of two 13 layer CNN networks, trained on the RSNA dataset. The architecture of the convolutional neural networks has a consistent pattern starting with two convolutional layers and then a max pool layer. The number of channels in the convolutional layers increased by twice the previous amount every two convolutional layers. At the end of the architecture there is a dropout layer to reduce the overfitting during training. Motivation for ensemble is to force the model to pay attention at relevant areas on CXR.

This method of attention takes the 512x512 image and takes random samples of 256x256 and feeds these samples as the input to the first CNN model to encourage the model to look at the important areas of the image. After training, this model is used to generate a 17x17 heatmap, by taking an array of 256x256 images from the 512x512 image. The heatmap is an important part of this method of attention and provides a lot of valuable information. These predictions are binary and turn into the heatmap of binary values. This gives the model an idea of where pneumonia could be, similar to how radiologists and doctors may be looking at the image. A second model is trained with the heatmaps and 256x256 CXR image. The heatmaps go through one channel of 17x17 convolutional layer and a dropout layer before being concatenated with the other path that takes the normal 256x256 images as input. The normal resized 256x256 images go through a convolutional neural network of similar architecture, also including 13 convolutional layers of the same pattern. They both go through one last dense layer to get a weighted average of these two outputs. The models are trained with a batch size of 16 and a learning rate which starts at 0.00001. Every 50 epochs the learning rate is multiplied by 0.9. Each of the convolutional neural networks train through the Adam Optimizer (Brownlee, 2019).

## 3. Data

### 3.1. Training

To train CNN models I used the RSNA dataset released by Kaggle in 2018. This dataset contains 26,684 frontal-view X-ray images. This dataset was split into 80% train and 20% test. After data augmentation on the training dataset, it contained 33463 images. This augmented dataset was split into 80% train and 20% validation. Each of these images was labeled with binary numbers, 0 being no pneumonia and 1 meaning the image has pneumonia. The images in the dataset are well balanced since the set consisted of images of both genders, all ages from young to old, as well as location of pneumonia ranging form the apex of the lung to the base of the lung. Furthermore, each of these images with pneumonia had an additional bounding box(es) value which gave coordinates to where the pneumonia was located in the image.

Each of these images was in the size of 1024x1024. I made copies of these images resizing them into 512x512

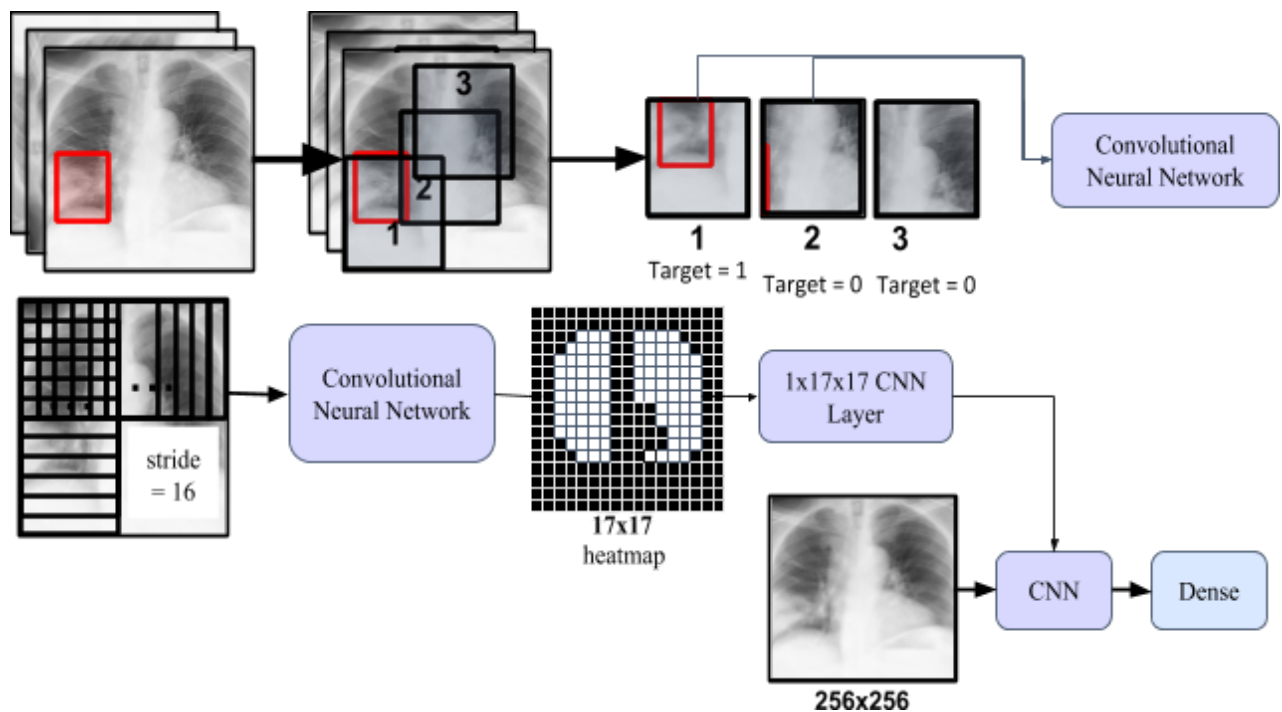

*Figure 2*: This diagram shows the steps and procedure through the unique method of attention I used to keep an emphasis towards the compensation of human fallibility. The goal of the method was to force the model to pay more attention to important areas of the image such as near the apex or behind the heart.

and 256x256. After experimenting with many different methods, my best method involved heat maps which were generated from a trained model on 256x256 patches in 512x512 images. The target for each of these patches was decided by the percentage of total pneumonia visible in the patch. I used a threshold of 10% to decide if the patch has pneumonia or not. I generated a heatmap of binary values by rolling a 256x256 window on each 512x512 image and predicting whether or not that image has pneumonia through the first model. The normal 256x256 image and 17x17 heatmap go through a second network. (shown in *figure 2*) This part of the attention method was done so that the model has a clear idea of where and what pneumonia looks like. This convolutional neural network acts differently than the first model which produced the heatmaps since it looks at the whole image as well as a heatmap providing possible areas of pneumonia instead of looking at only portions of the images. This allows the model to be more concise in its predictions. This method of attention provides the model with relevant information about possible regions to detect pneumonia through sampling of high-resolution images.

### 3.2. Test

The test set contained 5337, 256x256 frontal-view X-ray images. No data augmentation was done on the test set to make it as realistic as possible. Since the data had many more false images than positive images, the metrics I used to analyze our results were the F1 metric, recall, precision, and AUROC (Brownlee, 2019). I also took a random set of 25 images which were diagnosed by both the model and Dr. Sussane Soin, an expert radiologist. These 25 images were taken from both the NIH and RSNA datasets based on unique properties. To make

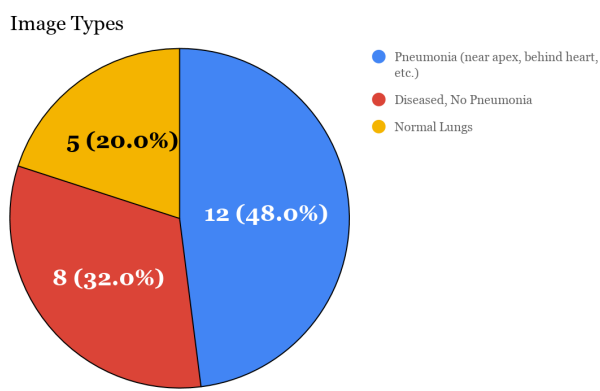

*Graph 1*: When choosing the 25 images to test and compare with humans and machines, images were chosen based on their unique properties. 12 images with pneumonia were chosen with some images where pneumonia appeared near the apex and behind the heart. 8 were chosen without pneumonia, but with other lung diseases. 5 normal images were chosen as well.

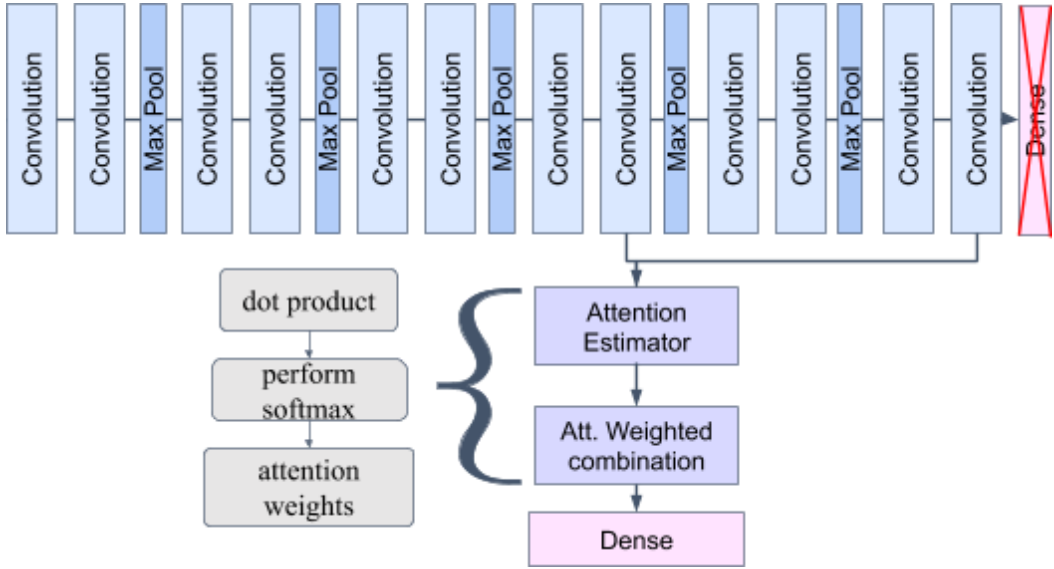

*Figure 3*: Soft attention is performed by taking two layers, the last layer and a layer of your choice to compare how similar the outputs are. Attention is given by enhancing the regions in the attended layer which are more similar to the final layer.

these results of greater value I choose reasonable images to test my model on (*graph 1)*.

## 4. PneumoXttention Vs. Attention Mechanism

### 4.1. Attention Mechanism

Attention first came out in 2017 in the paper, "Attention is all you need!", (Vaswani et al., 2017). They performed the attention mechanism in RNN. Later on, the attention mechanism was performed on CNN models, such as in the paper, "Learn to Pay Attention" (Jetley, Lord, Lee, & Torr, 2018). It demonstrates an approach towards soft attention, as demonstrated in the diagram above (*figure 3*). They gave an example of the implementation of a basic attention mechanism in a CNN through the basic formula of attention of a query, value, and key. I experimented with the placement of the attention mechanism and how it should be used during the training process. Some of the experiments consisted of choosing which layer to attend to and the number of attention layers we attended to in the model.

### 4.2. PneumoXttention Methodology

I tried the attention method as described in the paper "Learn to Pay Attention", however it did not improve the results. I wanted to improve the model by forcing it to focus on the parts of the image where pneumonia was located. This would make the final predictions of the model more reliable and trustworthy.

When analyzing the outputs of different models without attention, I noticed that often the model would not be looking in the right spots to detect the pneumonia. In fact, it would often be looking at boundaries or places outside the lung. (*figure 4*) This will not generalize and hence will have bad results on real images. For this reason, attention was necessary so that the model accurately detects pneumonia by looking at the correct places.

It can be given that with infinite images the model will be able to learn to pay attention to relevant areas of the image. To achieve the same effect, I randomly sampled 512x512 images to get a 256x256 patch. This allowed the model to learn to pay attention to important regions such as behind the heart and near the apex.

The second model was trained based on the previous models' predictions across the images creating a 17x17 heatmap. These heat maps provided useful information as to whether pneumonia was located on the image and where. The final portion of PneumoXttention was to use these heatmaps and the original 256x256 resized image as input to another convolutional neural network.

### 4.3. Comparison

There have been many algorithms other than PneumoXttention which have been trained to detect pneumonia. Algorithms like these include CheXNet, Wang et al, and Yao et al. All of these models were trained on the NIH dataset. Though these specific models were trained to detect 14 different diseases they have a separate AUROC for each disease. Furthermore, the model, CheXNet, main goal was to be able to accurately diagnose pneumonia to the best of its ability (Rajpurkar et al., 2017). CheXNet combined techniques from both Wang et al and Yao et al. to create a superior model.

Each model recorded the performance of the model for each disease as an AUROC. Wang et al. was the first released model which reached an AUROC of 0.633 (Wang et al., 2017). Yao et al. was released soon after reaching an AUROC of 0.713 (Yao et al., 2018). Not long after, CheXNet combined techniques used in both previous models and reached an AUROC of 0.768. My model, PneumoXttention, reached an AUROC of 0.95 when trained with this new method of attention. Without the method of attention applied to the original architecture of the model, the model would reach an AUROC of 0.76.

|  | PneumoXttention | Wang et al | Yao et al | CheXnet |
|---|---|---|---|---|
| ROC | 0.95 | 0.633 | 0.713 | 0.768 |

*Table 2*: PneumoXttention outperforms the top three networks at diagnosing pneumonia.

This shows that PneumoXttention performs significantly better when detecting pneumonia. (*table 2*)

## 5. Results

### 5.1. Metrics

To measure the performance of this algorithm I decided to use the f1 metric, recall, precision, and area under the ROC curve. The test set was imbalanced in the sense that there were more images whose target value was 0 and fewer images with a target value of 1. The recall is the number of positive images the model correctly identified divided by the total number of that have pneumonia. The precision is the number of positive images accurately classified by the model over the number of total images predicted true or with a value of one. The f1 score is the harmonic mean of precision and recall. The AUROC is the receiving operating characteristic curve which is often used to evaluate classification models.

### 5.2. Final Results

The model had a final f1 score of 0.82, a recall of 0.8, a precision of 0.84, and AUROC of 0.95 on the test set. (*table 3*)

| Precision | Recall | F1 score | AUROC |
|---|---|---|---|
| 0.84 | 0.8 | 0.82 | 0.95 |

*Table 3*: This table shows the final output metrics of PneumoXttention when predicting on the test set.

On the random 25 set of images, the model had an accuracy of 92% and the human had an accuracy of 72%. When combined the model completely compensated for human error meaning working together they could achieve 100% accuracy across all 25 images. Taking into account that radiologists look at more than the CXR image to diagnose patients of pneumonia, this was a successful model. These results showed that across these 25 difficult images the model accurately diagnosed every image the radiologist had gotten incorrect. This pattern was displayed on the opposite side as well. The radiologist accurately diagnosed every image the model incorrectly diagnosed.

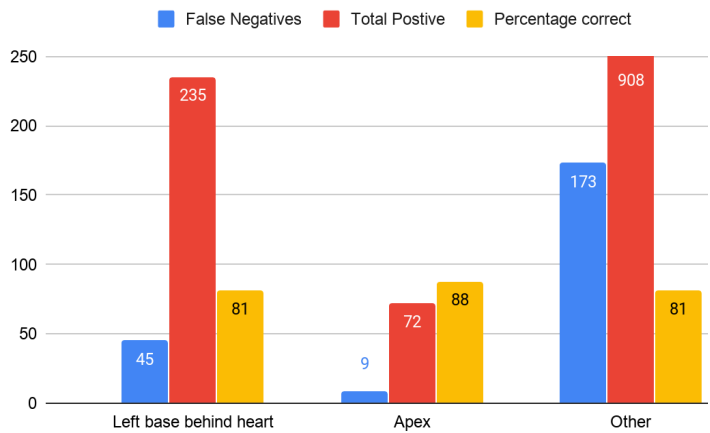

*Graph 2*: An analysis was performed to see how well the model performs when pneumonia appears in different regions of the image. Specifically, the left base behind the heart, near or around the apex, and everything else.

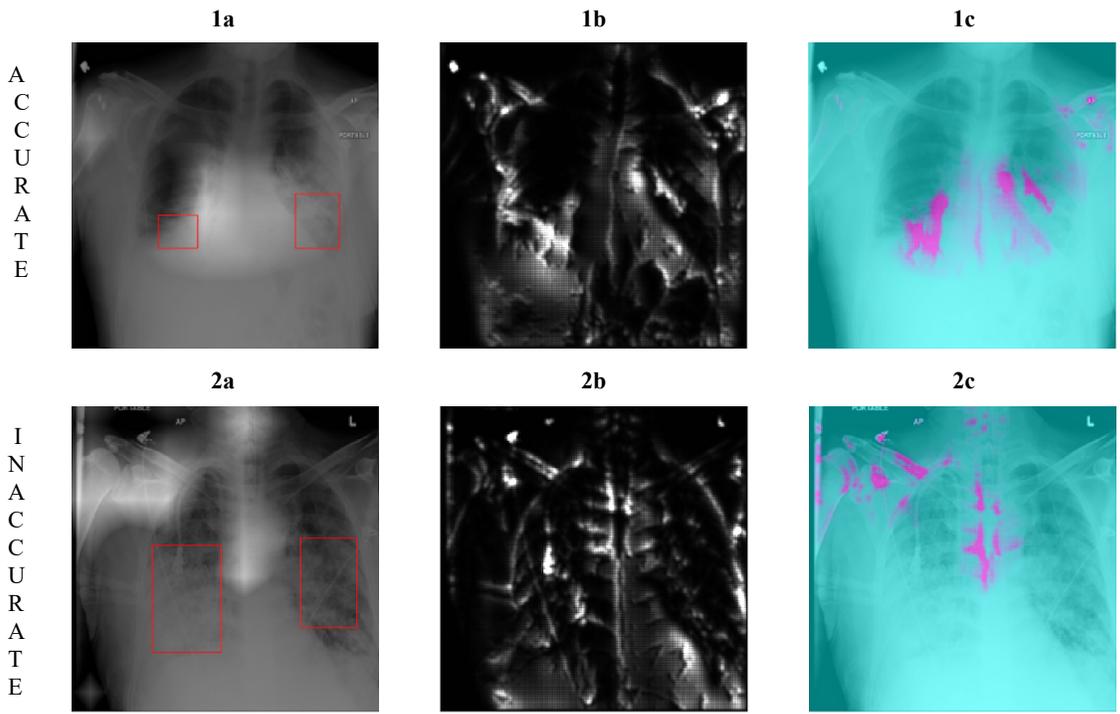

*Figure 4*: In picture (1a-1c) pneumonia is correctly diagnosed, correctly localized after attention
In picture (2a-2c) pneumonia is correctly diagnosed, incorrectly localized (bone structure) before attention.

The goal of the model was to compensate for human error, but PneumoXttention has an added benefit. It has the same accuracy of 81% in all cases no matter if pneumonia is near the apex or behind the heart (*graph 2*). Since human radiologists make more mistakes in these cases, PneumoXttention provides valuable additive information to the human Radiologist.

## 5.3. Analysis

Once PneumoXttention was trained, I tested it across the test set, constituting around 5337 images. When analyzing these images I found that the model was performing at an accuracy of 84.5% on images where pneumonia appears in "review areas", specifically behind the heart and neart the apex. It can also be noted that the model performed at a percentage of around 82% in all areas, including those which can often be inaccurately diagnosed by

| Image | Value | Human Diagnosis | m13h diagnosis |
|---|---|---|---|
| 1 | 0 | 1 | 0 |
| 2 | 1 | 1 | 0 |
| 3 | 0 | 0 | 0 |
| 4 | 1 | 1 | 1 |
| 5 | 0 | 1 | 0 |
| 6 | 1 | 0 | 1 |
| 7 | 1 | 1 | 1 |
| 8 | 0 | 0 | 0 |
| 9 | 0 | 0 | 0 |
| 10 | 1 | 1 | 1 |
| 11 | 1 | 1 | 1 |
| 12 | 0 | 0 | 0 |
| 13 | 0 | 0 | 0 |
| 14 | 0 | 0 | 0 |
| 15 | 0 | 0 | 0 |
| 16 | 1 | 0 | 1 |
| 17 | 1 | 0 | 1 |
| 18 | 1 | 1 | 1 |
| 19 | 0 | 0 | 0 |
| 20 | 0 | 0 | 1 |
| 21 | 1 | 1 | 1 |
| 22 | 1 | 0 | 1 |
| 23 | 0 | 0 | 0 |
| 24 | 1 | 0 | 1 |
| 25 | 0 | 0 | 0 |

*Table 4*: This table shows the 25 images the model and radiologist predicted. Green means correct whereas red means wrong.

radiologists. (*graph 2*)

To further examine the usefulness and performance of PneumoXttention the results were analyzed across a specific set of 25 images, which were selected based on specific properties *(described in section 3.2 Test)*.

To visually see the model accomplishing its goal, I analyzed the predictions between the final attention model and the radiologist's predictions (*table 4*), I asked Dr. Sussane Soin to analyze and predict whether each image had pneumonia or not. After comparing the results, it was noted that the model completely compensated for human error. Taking in mind the idea that radiologists base their final results on more than the chest radiograph images, this would only result in more accurate final results. Though this was only across 25 images, these were images both the model and radiologist had never seen before, meaning there was a high chance that this happens for a larger sample of images.(*figure 5*)

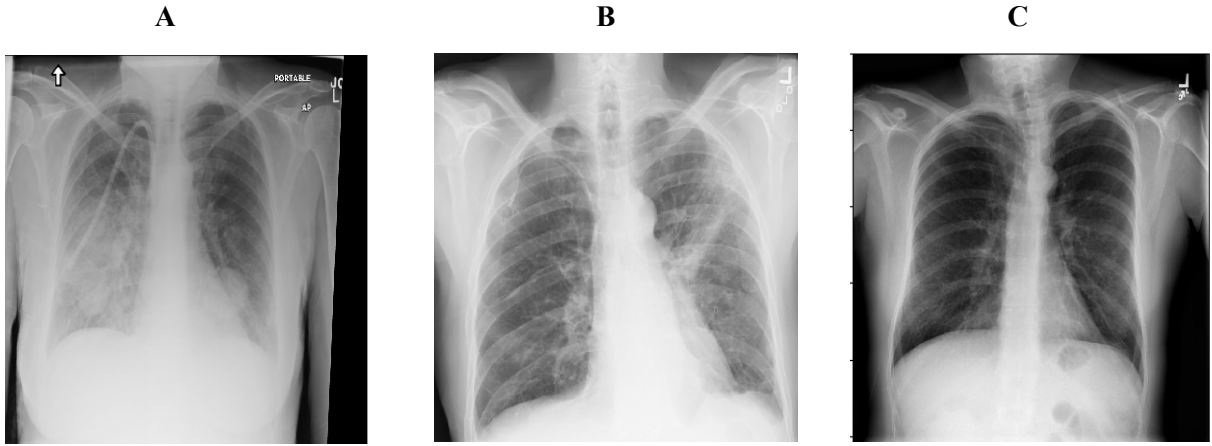

*Figure 5*: These images are a few of the images which were tested by both human and model.
**A:** Wrong Diagnosis by human, Correctly Diagnosed by model. (X-ray can be revisited for final diagnosis)
**B:** Correctly Diagnosed by human, Correctly Diagnosed by model. (Higher confidence in final diagnosis)
**C:** Correctly Diagnosed by human, Wrong Diagnosis by model. (X-ray can be revisited for final diagnosis)

## 6. Related Work

In many research universities and other facilities there have been recent advancements in the field of machine learning, to further improve the medical performance of AI through diagnosis, data analysis, etc. Diagnosis through chest radiograph x rays has always been quite popular in this field. Many improvements to this has been done by introducing new layers, advanced concepts, and better convolutional neural networks.

Kaggle released this RSNA dataset in order to inform and persuade people to take on this problem and help make the world a better, safer place to live.

Some of the papers released had the objective of creating a model which could detect pneumonia accurately, A few of these include models like CheXNet, Wang et al, and Yao et al.

With machines and models which can detect diseases and help doctors in real life, we can save more people both emotionally and physically. Millions of patients lose their lives due to disease and misdiagnosis. We can lift some of the weight and pressure of the healthcare personals' shoulders by giving them better tools at this task.

## 7. Conclusions

Pneumonia is a cause of many deaths in young children and older adults. Early, accurate diagnosis and treatment can be a big part of the solution to this problem. Although radiologists base their final diagnosis on more than the CXR image, it is an important piece in their diagnosis. In developing countries, there is a shortage of expert radiologists making the misdiagnosis more common. Tools like PneumoXttention can help in providing a second opinion to radiologists and hence improving overall diagnosis and saving lives. PneumoXttention was able to diagnose with 92 % accuracy while the human radiologist had an accuracy of 72% which can be derived from (*table 4*), and working together achieved 100% accurate results. In order to reduce the number of deaths that occur each year tools like PneumoXttention are useful.

# 8. Acknowledgments

Many people encouraged, helped and assisted me throughout this project. I want to start by thanking Dr. Sussane Soin, a consultant radiologist at London North West University Healthcare NHS Trust. She provided me with valuable feedback on this project as well as diagnosed 25 CXR test images that are used for testing models. I would also like to thank Dr. Colin Whitby-Strevens, a retired Apple Systems Engineer, who also provided me with much feedback and articles to encourage me to continue this project.

My science teacher Miss Ragu supported me throughout the project by organizing weekly science club meetings, where I discussed and researched topics related to my project. She introduced me to Professor, Dr. Colin Whitby Strevens, who gave me continued feedback. I am very grateful to her for wonderful teaching and all the support in my Science Fair Journey.

Finally, I would like to thank my father, Mr. Manish Singh, a Principal Design Engineer at Ambarella, who assisted me through this project, providing lots of help with coding and AI related problems. He was always there for me during difficult situations and sat down with me when I was confused or stressed. Finally, I would like to thank my mother, Kirti Singh, and younger sister, Manasvini Singh, who took their time to listen to all of my rambling and proofreading my write-ups.

# 9. Reproducing the Results

To access the code and reproduce the results visit: https://github.com/Sanskriti-Slngh/PneumoXttention.git

If you have any questions regarding this paper please email *sanskritisingh0914@gmail.com*.